\documentclass[12pt]{iopart}
\usepackage{graphicx}
\usepackage{graphics}
\usepackage{subfigure}
\usepackage{multirow}
\usepackage{wrapfig}
\usepackage{float}
\usepackage{cite}
\usepackage{amssymb}
\usepackage{calrsfs}

\begin{document}
\title{Self-consistent kinetic simulations of lower hybrid drift instability resulting in electron current driven by fusion products in tokamak plasmas}
\author{J W S Cook$^{1}$, S C Chapman$^{1}$ and R O Dendy$^{2,1}$}
\address{$^{1}$Centre for Fusion, Space and Astrophysics, Department of Physics, Warwick University, Coventry CV4 7AL, U.K.}
\address{$^{2}$Euratom/CCFE Fusion Association, Culham Science Centre, Abingdon, Oxfordshire OX14 3DB, U.K.}

\begin{abstract}
We present particle-in-cell (PIC) simulations of minority energetic protons in deuterium plasmas, which demonstrate a collective instability responsible for emission near the lower hybrid frequency and its harmonics. The simulations capture the lower hybrid drift instability in a regime relevant to tokamak fusion plasmas, and show further that the excited electromagnetic fields collectively and collisionlessly couple free energy from the protons to directed electron motion. This results in an asymmetric tail antiparallel to the magnetic field. We focus on obliquely propagating modes under conditions approximating the outer mid-plane edge in a large tokamak, through which there pass confined centrally born fusion products on banana orbits that have large radial excursions. A fully self-consistent electromagnetic relativistic PIC code representing all vector field quantities and particle velocities in three dimensions as functions of a single spatial dimension is used to model this situation, by evolving the initial antiparallel travelling ring-beam distribution of 3MeV protons in a background 10keV Maxwellian deuterium plasma with realistic ion-electron mass ratio. The simulations thus demonstrate a key building block of alpha channelling scenarios for burning fusion plasmas in tokamaks.

\end{abstract}

\maketitle

\section{Introduction}

There is sustained interest in the collective instabilities of energetic ion populations in plasmas with confining magnetic fields that involve emission and absorption of radio frequency (RF) waves. For fusion plasmas, the primary aspects of practical interest include the suprathermal ion cyclotron emission arising from the products of deuterium and tritium fusion reactions, observed in both JET and TFTR, typically in the $>$10 MHz range \cite{ref:dendy1995nuclfusion}; and, at the highest frequencies where electron dynamics also enter, alpha channelling\cite{ref:fisch1992prl}. This refers to the possibility of collectively coupling energetic ion excited RF wave energy to create useful distortions of particle velocity distributions and hence, for example, to efficiently drive electron plasma current. In the present paper we focus on the second of these topics. As noted in the pioneering theoretical work of Fisch and co-workers \cite{ref:fisch1992prl}, understanding and exploiting the physics of lower hybrid waves that are excited collectively by the energetic ions and subsequently absorbed by Landau damping on the electrons, is central to several aspects of alpha channelling. This raises interesting theoretical challenges: in particular, the need to address simultaneously and self consistently the collective dynamics of ions, for which the highest natural frequency is the ion cyclotron frequency $\Omega_{ci}$ (and its harmonics), and the dynamics of waves involving collective motion of the electrons, for which the lowest natural frequency is the lower hybrid frequency \cite{ref:stix}, $\omega_{LH}$, defined by

\begin{equation}
\omega_{LH}=\left({\frac{\Omega_{ce}\Omega_{ci}}{1+\Omega_{ce}\Omega_{ci}/\omega_{pi}^2}}\right)^{1/2}
\end{equation}

\noindent Here $\Omega_{ce}$, $\Omega_{ci}$ and $\omega_{pi}$ represent the electron and ion cyclotron frequencies and the ion plasma frequency respectively. Fusion alpha-particles are born at energies of 3.5MeV in deuterium-tritium fusion plasmas whose thermal energies are of order 10keV. This creates the possibility of population inversion in velocity space, at least transiently during burn initiation before the alpha-particles have slowed through collisions with electrons. Population inversion opens the door to rapid energy exchange through collective instability. It is already known that, in tokamak plasmas, spatially localised inversions of the energy distribution of fusion-born ions can be sustained for long periods as a result of the interplay between particle energy and the pitch angle-dependent character of particle drift orbits. Specifically, ion cyclotron emission was excited in the outer mid-plane edge region of JET and TFTR as a consequence of the distinctive radial excursions \cite{ref:cottrell1988prl,ref:cauffman1995nuclfusion} of the drift orbits of fusion products born with pitch angles just inside the trapped-passing boundary. This was first seen for fusion protons born in pure deuterium plasmas in JET \cite{ref:cottrell1988prl,ref:Schild1989nuclfus}, and subsequently for fusion alpha-particles in deuterium-tritium plasmas in JET \cite{ref:cottrell1993nuclfusion,ref:mcclements1999prl} and TFTR \cite{ref:dendy1995nuclfusion,ref:cauffman1995nuclfusion,ref:mcclements1996physplasmas}, as well as injected beam ions \cite{ref:dendy1994physplasmas} in TFTR and heated minority ions in JET \cite{ref:mcclements1999prl}. For reviews and further theory, we refer to \cite{ref:dendy1995nuclfusion,ref:dendy1994ppcf,ref:dendy1992physfluids}. These observations motivate the choice of model distribution for the energetic ions in the present study. Given the variety of possible alpha channelling physics scenarios, it is necessary to find a focus for the representation of population inversion in the present study, so we confine our attention to the primary type of ion distribution that has so far been observed to excite RF waves by collective instability in large tokamak edge plasmas. Specifically, we follow \cite{ref:dendy1992physfluids,ref:dendy1994ppcf,ref:mcclements1996physplasmas,ref:dendy1994physplasmas2} in adopting a simple model in which the distribution of energies perpendicular to the local magnetic field is narrow, and there is a parallel drift. We note that similar distributions, giving rise to similar collective instability leading to excitation of RF waves, have been observed in the terrestrial magnetosphere \cite{ref:mcclements1993jgr,ref:mcclements1994jgr}, and may be operative in the interstellar medium \cite{ref:mcclements1996mnras,ref:mcclements1997mnras,ref:kirk2001jpg}. Our choice of parameter values for the model energetic ion distributions, and for the majority thermal ion and electron populations that comprise the background plasma, is similarly motivated by the observations of ion cyclotron emission from large tokamak edge plasmas.

In this paper, we use a fully kinetic particle-in-cell (PIC) code to examine the collective excitation of waves near the  lower hybrid frequency by such energetic ion populations in tokamak edge plasma conditions. Electrons damp the waves produced by the instability and are thereby accelerated. This results in an asymmetric tail in the distribution of electron velocities in the direction parallel to the magnetic field, implying an electron current. The fully kinetic PIC simulation self-consistently evolves in time all field vector quantities and particle velocities in three dimensions as a function of one space coordinate for three populations $-$ energetic minority ions, thermal majority ions, and thermal electrons represented with realistic mass ratios. The results included in this paper expand and elaborate upon the mechanism identified in Ref. \cite{ref:jwscook2010arxiv} for the generation of a suprathermal tail in the distribution of electron velocities parallel to the magnetic field and consequently provides a building block for the construction of tokamak-relevant alpha channelling scenarios.

The linear phase of the instability underlying the phenomena seen in these fully nonlinear numerical simulations is consistent with a lower hybrid drift instability (LHDI). To simplify and generalise somewhat, the wave excited is predominantly electrostatic and propagates nearly perpendicular to the magnetic field, with a parallel phase velocity which (in the cases we consider) is sufficiently close to the electron thermal velocity to enable significant electron Landau damping. The wave frequency is high in comparison to the ion cyclotron frequency, so that the action of the energetic ions on the timescale of relevance is predominantly that of a quasi-unmagnetised beam.

The scenario we consider differs fundamentally from that considered in much of the LHDI literature in one important aspect. Most previous work on LHDI has considered a free energy source residing in bulk ion drifts that are consistent with, and contribute to, the overall equilibrium of the plasma; notably, diamagnetic drifts. For the tokamak application, however, it is necessary to consider the LHDI of diffuse minority energetic ion populations that do not contribute to the plasma equilibrium. This is the case both for the edge localised population that appears to drive ion cyclotron emission, and for many alpha channelling scenarios. This degree of freedom arises from the interplay between particle energy and particle orbits in tokamak magnetic geometry, which ultimately derives from the combination of the particle toroidal velocity with the toroidal component of the magnetic vector potential in the canonical toroidal momentum. It can enable the local spatial generation of anomalous velocity space structures.

The LHDI is a phenomenon arising from drifting populations of ions, and its character is highly dependent on the values of key plasma parameters \cite{ref:yoon2003physplas,ref:yoon2008physplas,ref:hsia1979physfluids,ref:davidson1977physfluids}. LHDI is present in natural \cite{ref:dobe1999prl,ref:yoon1994physfluids,ref:drake1983physfluids} and laboratory plasmas \cite{ref:gladd1976plasphys,ref:carter2001prl}. Work has been done on LHDI with reference to magnetic reconnection in both laboratory plasmas \cite{ref:carter2001prl} and magnetospheric \cite{ref:bale2002geophys,ref:huba1978physfluids,ref:daughton2004prl} plasmas. Inhomogeneities drive particle drifts in plasmas which generate the LHDI in the cases considered in most of the literature \cite{ref:silveira2002physreve,ref:davidson1977physfluids,ref:daughton2004prl,ref:gwal1979physscrip,ref:gladd1976plasphys,ref:carter2001prl,ref:huba1978physfluids}, whereas the much larger drift relative to the ion thermal velocity considered here is a consequence of the energy released per particle in fusion reactions and of the geometry of the confining magnetic field in the tokamak. A significant number of independent plasma parameters must be specified in calculations of the properties of the LHDI for a particular application. The quantitative characterisation of the LHDI in the research literature is thus inevitably incomplete, and will likely remain so, because it is possible to explore only patchily the multi-dimensional parameter space. This situation also applies as regards the nonlinear development of the LHDI. Davidson \etal \cite{ref:davidson1977physfluids} consider the LHDI using imposed drifts from field inhomogeneities for the strictly perpendicular case. Hsia \etal \cite{ref:hsia1979physfluids} includes finite $k_{\parallel}$ in their analytical approach. The more recent study by Silveira \etal \cite{ref:silveira2002physreve} builds on these studies, and other important works too numerous to mention by name here. They present a unified formulation for instabilities in the lower hybrid range of frequencies due to field inhomogeneities. In these cases the LHDI nevertheless is excited by inhomogeneity induced drifts that are slower, in dimensionless terms, than the fusion product drifts considered in this paper. The pioneering computational study by Chen and Birdsall \cite{ref:chen1983physfluids} has cold fluid electrons. Further numerical work by Yoon \etal \cite{ref:yoon2008physplas} investigates the LHDI with gyro-averaged electrons and kinetic ions, while work by Daughton \etal \cite{ref:daughton2004prl} uses physical mass ratio kinetic particle simulations for $\nabla n$ and $\nabla B$ inhomogeneities which generate low drift velocities.

Finally we note that PIC codes have been used to address astrophysical problems that involve the same combination of effects that is considered here: beam-type minority ion distributions, resulting in RF wave excitation that leads on, through damping, to electron acceleration \cite{ref:dieckmann2000astronastro,ref:mcclements2001prl,ref:schmitz2002apj,ref:schmidtz2002apj}. PIC codes are also widely used to study other fundamental instabilities. Population inversion \cite{ref:lee1979physfluids}, kink instability \cite{ref:daughton2002physplas}, Weibel instability\cite{ref:kato2005physplas}, Kelvin-Helmholtz instability \cite{ref:theilhaber1989prl}, shocks \cite{ref:lee2004apj,ref:lee2005physplas} and wave-particle interactions \cite{ref:devine1995jgr,ref:oppenheim1999prl,ref:dieckmann1999physplas,ref:birch2001physplas,ref:birch2001grl} are among the many avenues investigated using PIC.

\section{Simulations}

The 1D3V PIC code epoch1d, which is based on PSC \cite{ref:psc}, represents the full distributions of all the species in the simulation, using computational macroparticles in three dimensional velocity space and in one configuration space dimension. These are evolved in time via the relativistic Lorentz force law in the absence of collisions. From this particle representation of the distribution function, the plasma moments (charge and current densities) are interpolated onto a fixed spatial grid in one dimension. From these, using the full Maxwell equations, the electromagnetic fields are advanced on the grid. The PIC algorithm thus self-consistently advances three dimensional vector fields and particle velocities in one spatial coordinate $(x)$ and time.

The spatial domain is split into $N_G$ cells of width $\Delta x$, each of which has a grid point at the centre. Cell size, which is uniform across the domain, is set at $1/10$ the electron Debye length and resolves the gyroradii of all species. The phase space density function $f$ of each species is represented by a large (682670) set of computational macroparticles which are initially distributed evenly in space amongst 2048 cells. Although all components of velocity are evolved, only the velocity component in the direction of the simulation domain causes the particles' position to change. Each velocity component of all particles contributes to the respective current component on the grid points in nearby cells. The field and particle advance timestep, ${\Delta t}$, divided by the cell size, ${\Delta x}$, resolves the speed of light, and thus resolves cyclotron frequencies. Periodic boundary conditions are used for particles and fields and any wavemodes present propagate parallel or antiparallel to the direction of the simulation domain.

The trajectories of background thermal electrons and deuterons, both of which have non-drifting Maxwellian distributions with temperature of 10keV, are evolved in time alongside the minority energetic protons (initially at 3MeV) which initially have a ring-beam velocity distribution modelled by

\begin{equation}
f_{p} = \frac{1}{2\pi v_r} \delta(v_{\vert\vert} - u) \delta(v_{\perp} - v_{r})
\label{eqn:falpha}
\end{equation}

\noindent see for example Dendy et al. \cite{ref:dendy1994physplasmas2} and McClements et al. \cite{ref:mcclements1996physplasmas}. Here $u$ is the velocity along the magnetic field and $v_r$ is the perpendicular velocity of the ring. As mentioned previously, this choice of model is motivated by multiple studies of ion cyclotron emission in JET and TFTR. For this reason the initial pitch angle of the ring distribution (angle between $u$ and $v_r$) is initialised at $135^\circ$ with respect to the background magnetic field at the start of the simulation. The initial velocity distributions of all the species in the simulation are uniform in space. The angle of inclination of the direction of spatial variation $x$, of our simulation to the initial background magnetic field is $\theta = 84^\circ$. We simulate a quasi-neutral plasma with an electron number density $n_e = 10^{18}m^{-3}$, and an applied magnetic field of $B = 3T$. The electron beta is $\beta_e = 0.03\%$. The ratio of the initial energetic ion speed $v_p(t\!=\!0) \equiv {(u^2 + v_r^2)}^{1/2}$, to the Alfv\'{e}n speed $V_A$, is $v_p(t\!=\!0)/V_A \simeq 0.51$. This set of model parameters is selected to enable us to address quasi-perpendicular propagating modes under conditions approximating the edge plasma of a large tokamak, subject to the necessary simplifications. It optimizes the grid size with respect to the gyroradii of the particles, so as to resolve the essential physics with reasonable computational resources. The gyroradius of the energetic protons, and the gyroradii at thermal speeds of electrons and deuterons, are resolved by 965, 1.8 and 111 grid cells respectively. The ratio of protons to deuterons $n_{p}/n_{d} = 10^{-2}$. The 3MeV ring of protons is not replenished as the distribution function evolves away from the initial configuration; there is no source term. These parameter values imply that the total energy of the energetic proton population is 1.7 times that of the background thermal deuterons and electrons combined. This ratio exceeds by a factor $\sim10$ the value anticipated for various energetic ion populations in next step fusion plasmas, see for example Table 1 of S. Putvinski (Nucl. Fus. $\bf{38}$, 1275 (1998)). This value is necessary to drive the instability on an acceptable timescale given finite computational resources, and brings this study closer to fusion relevant regimes than much of the LHDI literature.

\section{Flow of energy between particles and fields}

\begin{figure}[H]
  \begin{center}
\subfigure[]{\label{fig:Energy}\includegraphics[]{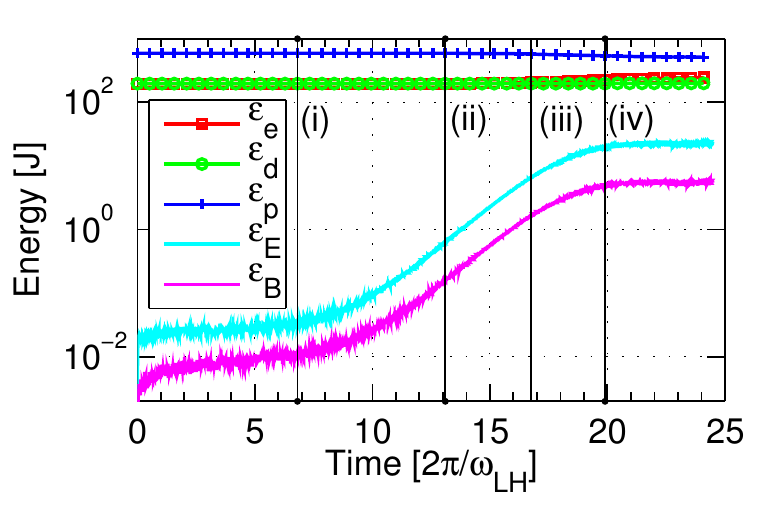}}
\subfigure[]{\label{fig:PartFluctEnergiesAndChangesInFieldEnergiesLog}\includegraphics[]{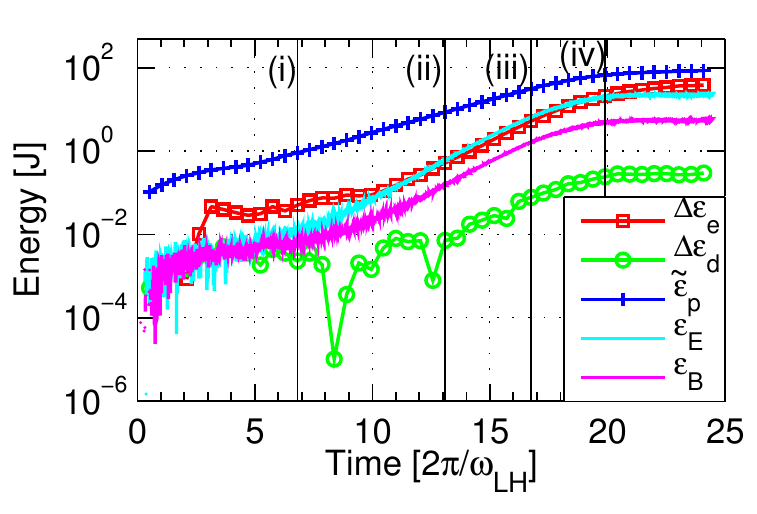}}
  \end{center}
\begin{small}
\caption{\label{fig:Energies}Panel (a): Time evolution of the kinetic energy, $\varepsilon_i$ defined by equation \ref{eqn:kineticenergy}, in each plasma species, and of the energy of electric field $\varepsilon_E$, and magnetic field $\varepsilon_B$ defined by equation \ref{eqn:electricfieldenergy}. The energy of the magnetic field excludes the applied component. Panel (b): Time evolution of the proton fluctuation energy, ${\tilde{\varepsilon}}_i$ defined by equation \ref{eqn:energyfluct}, change in kinetic energy of electrons and deuterons, $\Delta \varepsilon_{i} = \varepsilon_i(t) - \varepsilon_i(t=0)$, and of the electric and magnetic field energies $\varepsilon_E$ and $\varepsilon_B$. Both panels show vertical lines labelled $(i)\!-\!(iv)$ which denote the four snapshots in time at which particle data is shown in other figures. Time is measured in units of the lower hybrid period.}
\end{small}
\end{figure}

In order to examine the instability produced by the free energy in the energetic proton population we first consider the time evolution of the electric $\varepsilon_E$ and magnetic $\varepsilon_B$ total field energies in the simulation:

\begin{equation}
 \varepsilon_E=\frac{\varepsilon_0}{2}\sum_{i=1}^{N_G}E^2_i\Delta x \hspace{1cm}    \varepsilon_B=\frac{1}{2\mu_0}\sum_{i=1}^{N_G}(B^2_i - B^2_{0})\Delta x
\label{eqn:electricfieldenergy}
\end{equation}

\noindent where $B_{0}$ is the applied magnetic field. Figure \ref{fig:Energy} shows that the electric field energy, $\varepsilon_E$, (light blue trace) starts rising exponentially after a period of $\sim\!\!10\tau_{LH}$. The magnetic field energy, $\varepsilon_B$ (magenta trace), rises at approximately the same rate as the electric field, but contains nearly an order of magnitude less energy, implying that the waves produced by the instability are largely electrostatic. The total kinetic energy of species $i$ is defined by

\begin{equation}
 \varepsilon_i=w_i\sum_{j=1}^{N_{P,i}}(\sqrt{p_j^{2}c^2-m_i^2c^4}-m_ic^2)
\label{eqn:kineticenergy}
\end{equation}

\noindent where $w_i$ is the weighting factor to map the energy of each computational macroparticle, $j$, to the energy of the number of physical particles it represents. Particle weight is defined as $w_i = n_{0,i}L/N_{P,i}$ where $n_{0,i}$ is the species number density, $L$ is the length of the spatial domain, and $N_{P,i}$ is the number of computational macroparticles that represent the species in the simulation. The essential features of figure \ref{fig:Energy} indicate a linear phase of field growth during $10 \lesssim t/\tau_{LH} \lesssim 15$, accompanied by a corresponding small decline in total proton energy. Field amplitudes remain approximately constant after $t \gtrsim 20\tau_{LH}$, while proton energy continues to decline and is matched by an increase in electron energy.

We now define a quantity which better shows the variations in kinetic energy of the proton population, which we denote as fluctuation energy $\tilde{\varepsilon}$ summed over the $j=1-N_{P,i}$ computational macroparticles of each $i^{th}$ species:

\begin{equation}
 \tilde{\varepsilon}_{i}(t) = \sum_{j=1}^{N_{P,i}} (|\varepsilon_j(t) - <\!\varepsilon_i(t=0)\!>|)
\label{eqn:energyfluct}
\end{equation}

\noindent The quantity $\tilde{\varepsilon}$ is the sum of the modulus of the change in kinetic energy of each particle from the initial rest frame mean energy value of the species, $<\!\varepsilon_i(t=0)\!>$. Thus $\tilde{\varepsilon}_i$ captures kinetic energy dispersion regardless of sign, whereas a time evolving change in total kinetic energy, $\Delta \varepsilon_{i} = \varepsilon_i(t) - \varepsilon_i(t=0)$ where $\varepsilon_i(t=0) = \sum_{j=1}^{N_{P,i}}\varepsilon_j(t=0)$, only represents changes in the combined energy of the ensemble of particles. In particular we use $\tilde{\varepsilon}$ to quantify the initial energy dispersion of the protons during the early phases of the instability, see figure \ref{fig:PartFluctEnergiesAndChangesInFieldEnergiesLog} and the discussion in section \ref{sec:protonvspacedynamics}, because in some respects it is less susceptible to noise.

Figure \ref{fig:PartFluctEnergiesAndChangesInFieldEnergiesLog} shows in greater detail the species energy dynamics. The values of proton fluctuation energy $\tilde{\varepsilon}_p$, and of the changes $\Delta \varepsilon_e$ and $\Delta \varepsilon_d$ in total electron and deuterium kinetic energy are plotted alongside changes in $\varepsilon_E$ and $\varepsilon_B$. The value of $\tilde{\varepsilon}_p$ grows, ultimately increasing by three orders of magnitude, whereas figure \ref{fig:Energy} shows that the total proton kinetic energy declines by much less than one order of magnitude. This reflects the role of the proton population as the source of free energy. The change in electron kinetic energy $\Delta \varepsilon_e$ rises in close alignment with electric field energy $\varepsilon_E$ during the linear stage of the simulation, $10 < t/\tau_{LH} < 15$. This is evidence of the electron oscillation within the largely electrostatic waves excited by the instability. The corresponding effect for the deuterons is also visible in figure \ref{fig:PartFluctEnergiesAndChangesInFieldEnergiesLog} but is much less due to their higher mass.

\begin{figure}[H]
  \begin{center}
\subfigure[]{\label{fig:ProtonMomentumPDFt0}\includegraphics[]{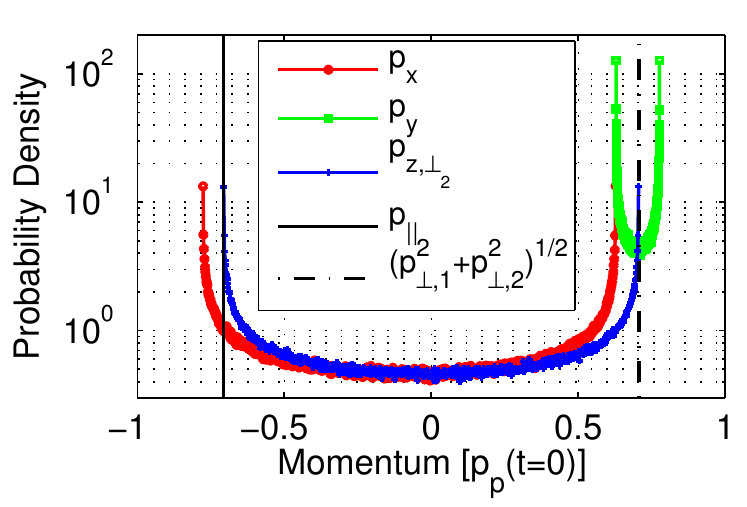}}
\subfigure[]{\label{fig:ElectonPparaPDF}\includegraphics[]{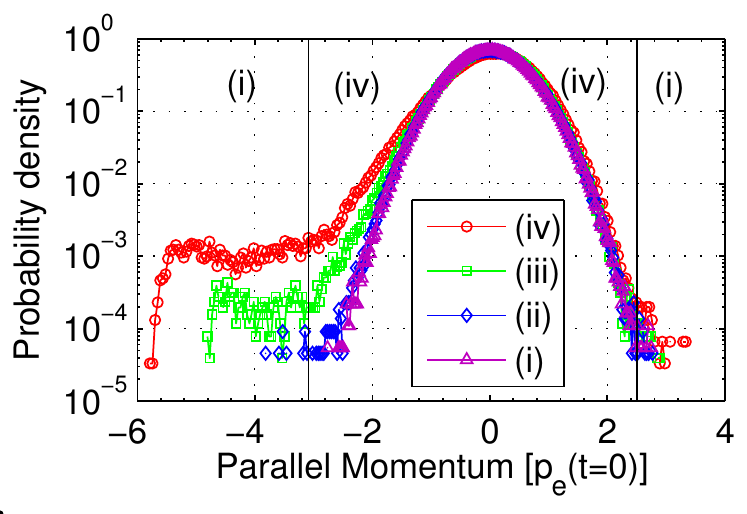}}
\subfigure[]{\label{fig:SkewnessAndEnergy3}\includegraphics[]{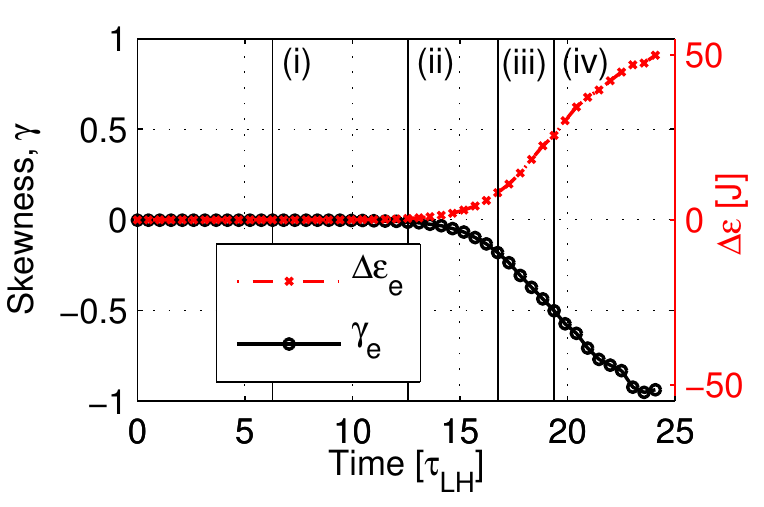}}
\subfigure[]{\label{fig:Currents}\includegraphics[]{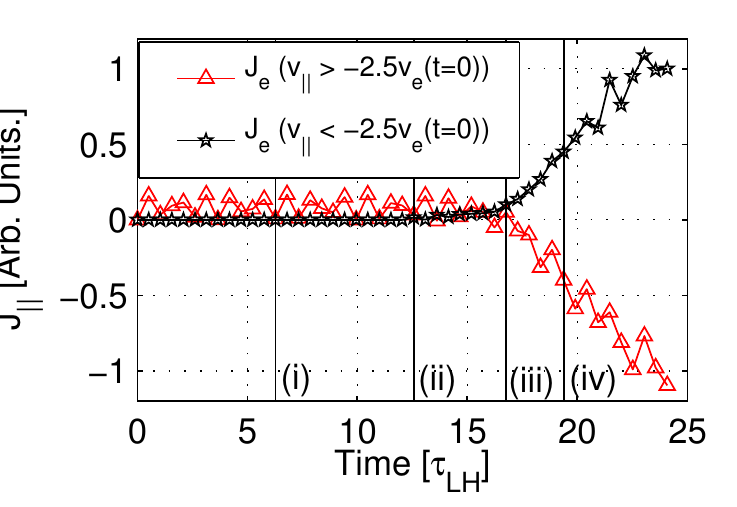}}
  \end{center}
\begin{small}
\caption{\label{fig:ElectonVparaPDFAndSkewnessAndEnergy}Panel $(a)$: Initial proton momentum PDF: along the simulation domain, $p_x$; in the plane created by the magnetic field and the simulation domain, $p_y$; perpendicular to both magnetic field and simulation domain $p_{z} \equiv p_{\perp_2}$; parallel to the magnetic field $p_{\parallel}$; and total perpendicular $p_{\perp} \equiv (p_{\perp,1}^2+p_{\perp,2}^2)^{1/2}$. Panel (b): Electron PDFs in relativistic parallel momentum $p_{\parallel}$, for the four snapshots in time $(i)\!-\!(iv)$ defined in Fig. \ref{fig:Energies}, also marked by vertical lines in panels $(c)$ and $(d)$. Vertical lines indicate the phase velocities of the forward and backward travelling dominant waves.  Panel (c): Skewness $\gamma_e$ of the electron $v_{\parallel}$ PDF (black circles) and change in electron kinetic energy $\Delta \varepsilon_e = \varepsilon_e(t) - \varepsilon_e(t=0)$, (red crosses). Panel (d): Time evolving parallel current in the electron bulk $v_x > -2.5v_e(t=0)$ (red triangles), and in the electron tail $v_x < -2.5v_e(t=0)$ (black stars). Momentum axes are in units of the initial relativistic momentum $p_i(t\!=\!0) = \left[(1/N_{P,i})\Sigma_{j=1}^{N_{P,i}}{\bf{p}}_j.{\bf{p}}_j(t=0)\right]^{1/2}$ of the species $i$, and time axes are in units of the lower hybrid oscillation period.}
\end{small}
\end{figure}

Figure \ref{fig:ProtonMomentumPDFt0} shows projections in momentum space of the PDF of the proton population at $t=0$ defined by (\ref{eqn:falpha}). The ring-beam distribution, as aligned with respect to the magnetic field, is projected into the cartesian co-ordinates of the simulation where $x$ is along the simulation domain, $z$ is perpendicular to both the simulation domain and the magnetic field and $y$ completes the orthogonal set. The initial PDF is a delta function in both $p_{\parallel}$ and $p_{\perp} \equiv (p_{\perp,1}^2+p_{\perp,2}^2)^{1/2}$. The three momentum components are projected into the simulation domain, in which the ring distribution becomes a more complicated function. Due to the oblique angle between simulation domain and applied magnetic field, the peaks in the PDF in $p_x$ (red circled trace) are not equal and opposite. This has consequences for the wave-particle interaction which are dealt with in the next section.

Figure \ref{fig:ElectonPparaPDF} shows electron parallel momentum at four different snapshots in time: $(i)\equiv6.8\tau_{LH}$ , $(ii)\equiv13.1\tau_{LH}$, $(iii)\equiv16.8\tau_{LH}$, $(iv)\equiv19.9\tau_{LH}$. Henceforth, these snapshots in time will be referred to by their Roman numeral.

Figures \ref{fig:ElectonPparaPDF} and \ref{fig:SkewnessAndEnergy3} show that for $t>15\tau_{LH}$ the electron distribution develops an asymmetric tail in $v_{\parallel}$ with corresponding skewness,

\begin{equation}
\gamma(t) = 1/N_{P,i}\Sigma_{j=1}^{N_{P,i}} \left(\frac{v_{\parallel,j}(t) - <{v_{\parallel,i}}(t)>}{\sigma_{v_{\parallel,i}}(t)}\right)^3
 \label{eqn:skewness}
\end{equation}

\noindent where $<\!{v_{\parallel,i}}(t)\!>$ and $\sigma_{v_{\parallel,i}}(t)$ are respectively the mean and standard deviation of electron $v_\parallel$ at time $t$.

Figure \ref{fig:Currents} shows the time evolving parallel electron current summed across configuration space. We plot the current for the bulk ($v_{\parallel} > -2.5v_e(t=0)$) and tail ($v_{\parallel} < -2.5v_e(t=0)$) of the electrons separately. We see that approximately equal and opposite currents are created by the bulk and tail electron populations in this collisionless simulation. This is a consequence of the periodic boundary conditions used in the code. The rate of change of current across the simulation domain under periodic boundary conditions follows from Amp\`{e}re's and Faraday's laws:

\begin{equation}
\mu_0 \frac{\partial}{\partial t} \int_0^L {\bf{J}} dx = - \int_0^L \nabla\!\times\!\nabla\!\times {\bf{E}} dx - \mu_0\epsilon_0 \frac{\partial^2}{\partial t^2} \int_0^L {\bf{E}} dx
\label{eqn:bcs1}
\end{equation}

\noindent where in one dimension ${\bf{\nabla}} = (\partial/\partial x ,0,0)$ and $\nabla\times\nabla\times \equiv (0,-\frac{\partial^2}{\partial x^2},-\frac{\partial^2}{\partial x^2})$.
Periodic boundary conditions enforce $F(0,t) = F(L,t)$ and $\nabla F(0,t) = \nabla F(L,t)$ where $F$ is any component of the electromagnetic field. Equation (\ref{eqn:bcs1}) is a wave equation in the plasma medium. For current density integrated over the box
\begin{eqnarray}
\left(c^2\nabla\!\!\times\!\!\nabla\!\!\times - \frac{\partial^2}{\partial t^2}\right) \int_0^L {\bf{J}} dx = 1/\epsilon_0 \frac{\partial}{\partial t} {\bf{\sigma}} \int_0^L {\bf{J}} dx
\end{eqnarray}
where ${\bf{J}} = {\bf{\sigma}}{\bf{E}}$ and ${\bf{\sigma}}$ is the plasma conductivity tensor. Consequently, any current density integrated across the domain may oscillate about zero but cannot grow in time for these simulations. We also note that collisions would differentially affect the bulk and tail electron populations, and hence their relative drift; however our PIC code does not incorporate collisions.

The asymmetry in the electron parallel PDF, reflecting net directional electron acceleration, continues to grow in the period beyond $t = 20\tau_{LH}$, during which wave energy quantified by $\varepsilon_E$ and $\varepsilon_B$ is approximately constant, see figure \ref{fig:Energy}. We infer Landau damping of the excited waves on resonant electrons, which results in the flattening of the electron parallel momentum PDF in figure \ref{fig:ElectonPparaPDF}. The vertical lines in this panel indicate the electron nonrelativistic momentum that corresponds to the phase velocity of the dominant relevant features in the $\omega,k$ transform of the electromagnetic field, to which we now turn.

\section{The nature of the electromagnetic fields}

\begin{figure}[H]
  \begin{center}
    \subfigure[]{\label{fig:DispRel_5lt_t_lt10}\includegraphics[]{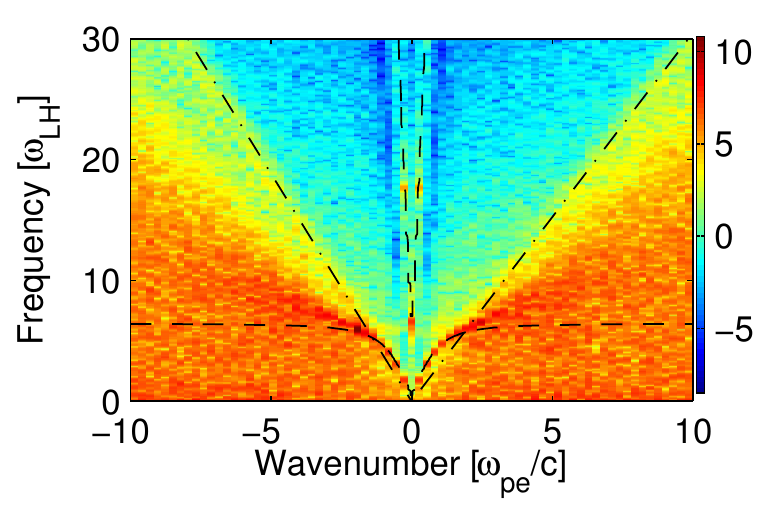}}
    \subfigure[]{\label{fig:DispRel_10lt_t_lt15}\includegraphics[]{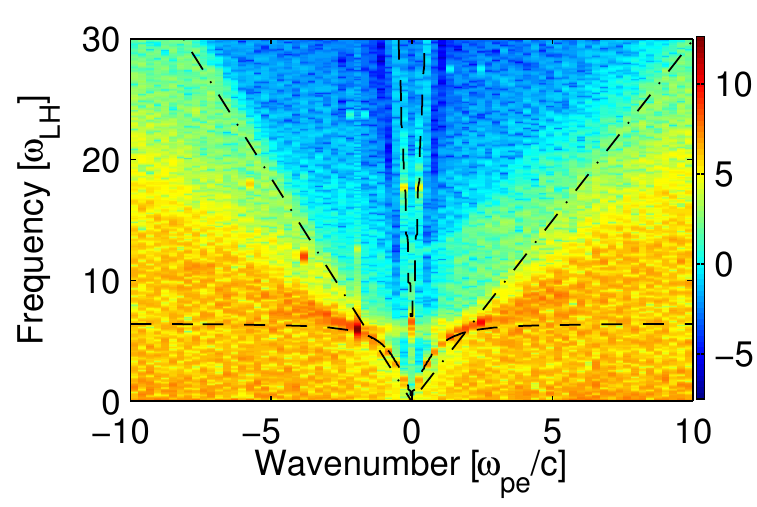}}
    \subfigure[]{\label{fig:DispRel_15lt_t_lt20}\includegraphics[]{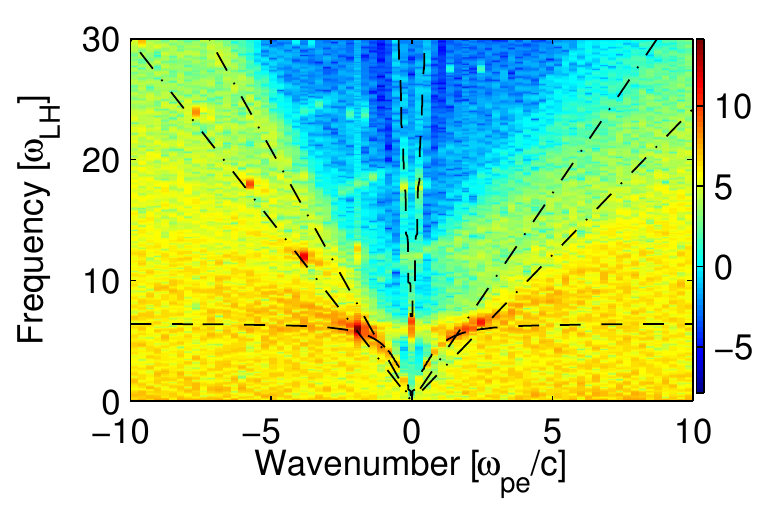}}
  \end{center}
\begin{small}
\caption{\label{fig:dispersionrelations}Spatio-temporal FFTs of the electric field component in the direction of the simulation domain $E_x$, integrated over all $x$ and the following time periods. Panel (a): $5 \leq t/\tau_{LH} \leq 10$. Panel (b): $10 \leq t/\tau_{LH} \leq 15$. Panel (c): $15 \leq t/\tau_{LH} \leq 20$. Overplotted on all panels is the cold plasma dispersion relation (dash trace). Overplotted on panel (a) and panel (b) are velocities corresponding to the peaks in the initial proton $v_x$ PDF (dot-dash trace). Overplotted on panel (c) are the velocities of the peaks  and extremes in the proton $v_x$ PDF at snapshot $(iv)$ (dot-dash trace). The frequency axes are in units of lower hybrid frequency and wavenumber axes in inverse electron skin depths $\omega_{pe}/c$.}
\end{small}
\end{figure}

Figure \ref{fig:dispersionrelations} displays the spatio-temporal fast Fourier transform (FFT) of the electric field component along the simulation domain $E_x$, for three epochs of the simulation, showing temporal evolution of the $\omega,k$ structure. Referring to figure \ref{fig:Energies}, the three epochs correspond to the following phases of electromagnetic field evolution: initial, $5 \leq t/\tau_{LH} \leq 10$; early linear growth, $10 \leq t/\tau_{LH} \leq 15$; and late linear growth, $15 \leq t/\tau_{LH} \leq 20$. The peaks in intensity at $k \simeq \pm 2 \omega_{pe}/c$ and $\omega \simeq 6\omega_{LH}$ in all three panels of figure \ref{fig:dispersionrelations} reflect the resonant interaction of the energetic initial proton population with the normal mode of the background plasma to which they most easily couple, namely the electron-deuteron lower extraordinary-Whistler wave mode whose analytical characteristics are outlined below.

Figure \ref{fig:DispRel_5lt_t_lt10} shows the waves present in the simulation during its initial phase before the onset of linear growth of electric and magnetic fields, in the time interval $5 < t/\tau_{LH} < 10$. Overplotted is the cold plasma electron-deuteron dispersion relation, implicitly defined by

\begin{equation}
 \left|\left(\begin{array}{ccc}
\varepsilon_1 -N^2 & -i\varepsilon_2 & 0\\
i\varepsilon_2  & \varepsilon_1 -N^2 cos^2(\theta)  & N^2 sin(\theta) cos(\theta)\\
0 & N^2 sin(\theta) cos(\theta) & \varepsilon_3 - N^2 sin^2(\theta)
      \end{array}
\right)\right| = 0
\label{eqn:dielectrictensor}
\end{equation}

This is the determinant of the cold plasma dielectric tensor for arbitrary propagation direction $\theta$ with respect to the magnetic field, where the refractive index $N = ck/\omega$. The components of the dielectric tensor are

\begin{eqnarray}
\varepsilon_1 = 1 + \frac{\omega_{pe}^2}{\Omega_{ce}^2 - \omega^2} + \frac{\omega_{pd}^2}{\Omega_{cd}^2 - \omega^2}\nonumber\\
\varepsilon_2 = \frac{\Omega_{ce}^2}{\omega^2}\frac{\omega_{pe}^2}{\Omega_{ce}^2 - \omega^2}
+\frac{\Omega_{cd}^2}{\omega^2}\frac{\omega_{pd}^2}{\Omega_{cd}^2 - \omega^2}  \nonumber\\
\varepsilon_3 = 1 - \frac{\omega_{pe}^2}{\omega^2} 
\end{eqnarray}

\noindent where $\omega_{pi}$ and $\Omega_{ci}$ are the plasma and cyclotron frequencies of species $i$. 

The cold plasma dispersion relation shows predominantly perpendicular propagating normal wavemodes; ordinary and extraordinary. However some properties of parallel propagating waves (Whistler) are additionally present, because the angle of wave propagation is oblique. The result is an extraordinary-Whistler branch with a cut-off frequency $> \omega_{LH}$. The instability is still essentially a lower hybrid drift-type instability, although it is active at frequencies above $\omega_{LH}$.

Figure \ref{fig:DispRel_10lt_t_lt15} is obtained using data from the interval $10 < t/\tau_{LH} < 15$. Figure \ref{fig:Energy} shows that this corresponds to the early linear stage of electric field wave growth, during which the changes in energy of electrons and deuterons also rise in tandem, reflecting their coherent participation in wave motion. The trace corresponding to snapshot $(ii)$ in figure \ref{fig:ElectonPparaPDF} shows that a monotonically decreasing electron tail is drawn out during this epoch, with skewness becoming apparent (figure \ref{fig:SkewnessAndEnergy3}) at its end. Figure \ref{fig:DispRel_10lt_t_lt15} enables us to identify the $\omega,k$ character of the waves excited in the linear growth phase of the simulation. Both forward and backward propagating waves are present, at $6\omega_{LH} \lesssim \omega \lesssim 8\omega_{LH}$ and $2\omega_{pe}/c \lesssim |k| \lesssim 5\omega_{pe}/c$. Figure \ref{fig:DispRel_10lt_t_lt15} also exhibits two peaks in electric field intensity which appear to be backward propagating harmonics of the fundamental activity at frequency $\omega \simeq 6\omega_{LH}$. During the late linear phase $15\leq t/\tau_{LH} \leq 20$, these multiple higher harmonics become established.

Figure \ref{fig:DispRel_15lt_t_lt20} shows data from $15 < t/\tau_{LH} < 20$ which encompasses the end of the linear growth phase during which electron acceleration arises (figures \ref{fig:ElectonPparaPDF} and \ref{fig:SkewnessAndEnergy3}). The dominant Fourier structures of the electric field are similar to the linear phase but with more pronounced backward propagating harmonics. Traces corresponding to snapshots $(iii)$ and $(iv)$ in figure \ref{fig:ElectonPparaPDF} show that the electron $p_{\parallel}$ PDF becomes increasingly flattened around the resonant momentum during this stage.

From figure \ref{fig:DispRel_10lt_t_lt15} we see that the peaks in energy in $\omega,k$ space correspond to the region around $\omega = 6\omega_{LH}$ and $k = \pm 2\omega_{pe}/c$. The precise values of $\omega,k$ of the dominant waves correspond approximately to where the phase velocity of the mode matches the velocities of the peaks in the proton $v_x$ PDF. Thus we can infer the parallel phase velocity of forward $v_+$, and backward $v_-$, travelling waves by reading off coordinates of the peaks in power from the $\omega,k$ plots. Thus the electrons that are in resonance with these waves have forward and backward parallel momenta of 

\begin{eqnarray}
p_{\parallel,+} = -v_{+}m_e/\cos\theta \simeq 2.5p_e(t\!=\!0) \nonumber\\
p_{\parallel,-} = -v_{-}m_e/\cos\theta \simeq -3.1p_e(t\!=\!0)
\label{eqn:}
\end{eqnarray}
\noindent in the nonrelativistic limit respectively. These momenta are indicated in figure \ref{fig:ElectonPparaPDF} by vertical lines. These show that the instability driven by energetic protons, for these simulation parameters, generates waves with phase velocity resonant with the tail of the initial Maxwellian distribution of electron velocities, where both the gradient and the small number of particles are conducive to efficient Landau damping. This gives rise to electron acceleration and a strongly asymmetric velocity tail: hence current drive.

\section{Proton velocity space dynamics}

\label{sec:protonvspacedynamics}
The particle-in-cell simulation presented here provides a full kinetic description of the species evolution in configuration space and velocity space. This differs from, and complements, analytical approaches. In the linear phase, analytical treatments are typically based on a small amplitude single harmonic perturbation. The linear phase here involves, as we have seen, fields mediating subtle interactions between the energetic protons, background deuterons, and electrons, in ways that would be difficult to capture in detail analytically. In the present section we therefore focus on the kinetics of the energetic proton population that drives the system, thereby obtaining a fresh perspective on the character of the instability during its lifetime.

\begin{figure}[H]
\begin{center}
\includegraphics[]{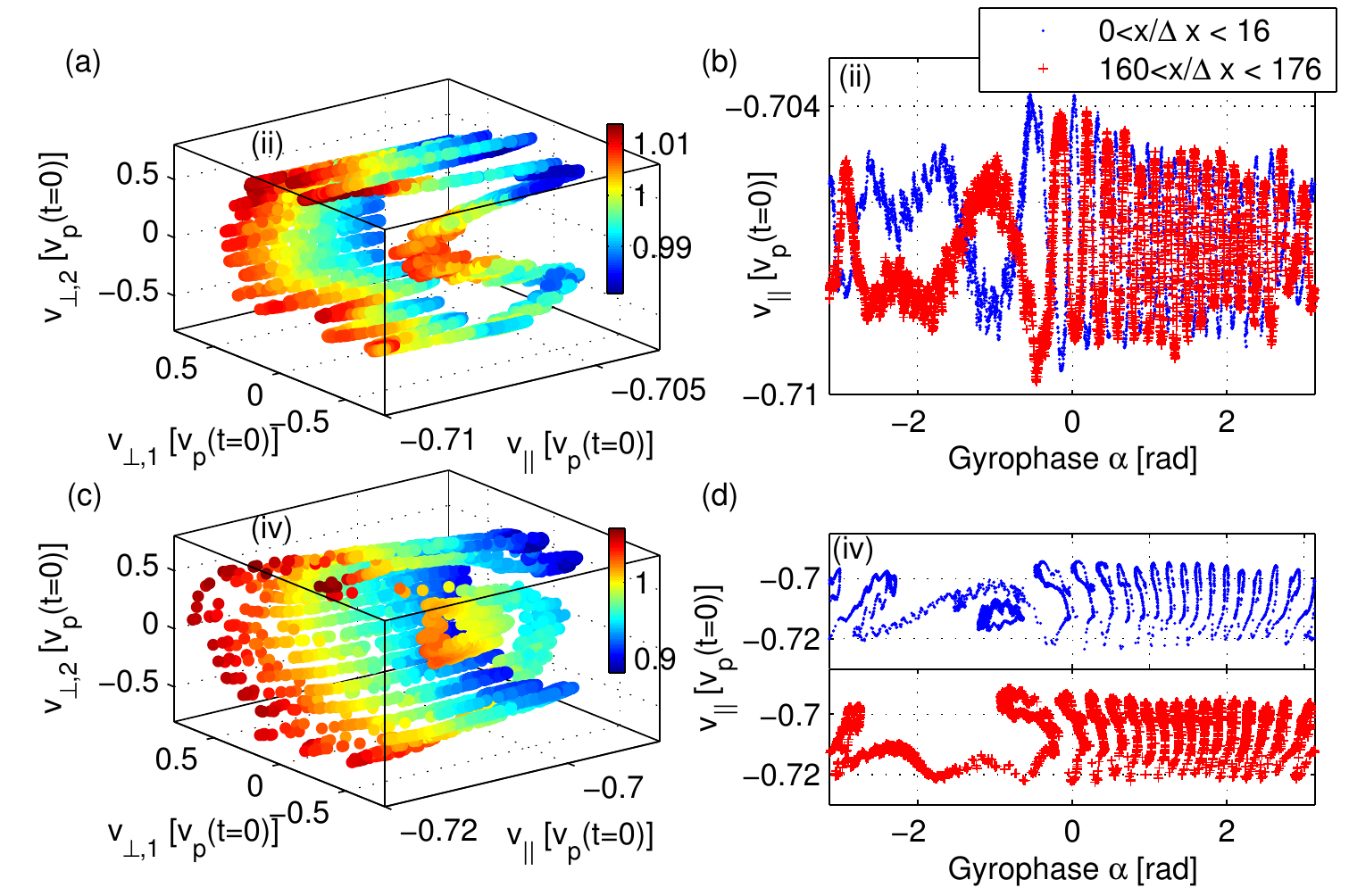}
\begin{small}
\caption{\label{fig:VspaceAndVparaVsgyrophase}Left panels: Velocities in magnetic field-aligned coordinates of protons found in the configuration space region $0 < x/\Delta x < 16$, where $\Delta x$ is the spatial extent of one grid cell. Colour indicates speed in units of $v_p(t=0)$. Right panels: Proton $v_{\parallel}$ as a function of gyrophase $\alpha = \arctan{v_{\perp,1}/v_{\perp,2}}$ for protons from two configuration space regions: $0 < x/\Delta x < 16$ (blue dots) and $160 < x/\Delta x < 176$ (red pluses). Top panels show data from snapshot ($ii$), bottom panels from snapshot $(iv)$. Velocity axes are in units of the proton characteristic velocity, $v_p(t=0) = \left[(1/N_{P,p})\Sigma_{j=1}^{N_{P,p}}{\bf{v}}_j.{\bf{v}}_j(t=0)\right]^{1/2}$, where $N_{P,p}$ is the number of computational macroparticles representing the protons.}
\end{small}
\end{center}
\end{figure}

Figure \ref{fig:VspaceAndVparaVsgyrophase} provides snapshots of the proton distributions in velocity space at two instants: $(ii)$, during the linear phase (upper panels); and $(iv)$, at the end of the linear phase (lower panels). The two left panels in figure \ref{fig:VspaceAndVparaVsgyrophase} focus on velocity space. They show the proton velocity distribution plotted with respect to three axes defined by $v_{\parallel}$ and two orthogonal velocity components $v_{\perp,1}$ and $v_{\perp,2}$ in the directions perpendicular to ${\bf{B}}_0$. Data plotted are for protons from a distinct region in configuration space as described below. The two right hand panels focus on configuration space behaviour. Each panel compares two subsets of protons. Each subset is localised in a distinct region of configuration space of extent $\delta x$, where $\delta x \ll \lambda$ and $\lambda = 2\pi/(2\omega_{pe}/c)$ is the wavelength of the dominant electromagnetic wave. The two distinct regions are separated in configuration space by $\sim\!\lambda/2$. To achieve this, protons are binned into subsets in configuration space such that the ratio of the size of each bin to the wavelength of the dominant electric field wave is $\delta x/\lambda \simeq 0.057$. The bins are separated by a gap of $10\delta x = 0.57\lambda$.

The upper panels of figure \ref{fig:VspaceAndVparaVsgyrophase} show early linear phase behaviour of the proton ring beam. There is a distinct wave-like structure oscillating in $v_{\parallel}$ around the gyro-orbit, shown in panel (b) for two subsets of protons. The two datasets show velocity space waves that are in anti-phase with each other, consistent with resonance with a wave at locations $\lambda/2$ apart.

The two bottom panels of figure \ref{fig:VspaceAndVparaVsgyrophase} show data from the end of the linear phase at snapshot $(iv)$. The proton $v_{\parallel}$ structure as a function of gyrophase from the same two configuration space bins as in panel (b) is shown for contrast in the panel (d). The waves in $v_{\parallel}$ are beginning to break nonlinearly.

The predominantly drift, as distinct from gyroresonant, character of the instability can be seen in the fully resolved velocity space of the protons. However it is challenging to extract the nature of the interaction between protons and the excited waves, because snapshots in time of proton phase space contain the history of all previous interactions. We now present a method for modelling the effect of previous interactions using only unperturbed proton trajectories which, of course, are easily calculable.

In this linear phase the interaction between the protons and the waves is weak: the protons are not trapped but follow trajectories through phase space close to those followed by gyro-orbits in the unperturbed fields. We can therefore directly verify that the large scale phase space structure in figure \ref{fig:VspaceAndVparaVsgyrophase}(b) arises from interaction with the dominant wave at $\omega \simeq 6\omega_{LH},k=-2\omega_{pe}/c$. To do this, we will reconstruct a phase space snapshot to compare with figure \ref{fig:VspaceAndVparaVsgyrophase}, assuming that: protons move on unperturbed gyro-orbits; and the perturbations in $v_{\parallel}$ are proportional to the wave amplitude experienced by the proton when last at resonance. Thus we can see the history of previous wave-particle interactions in the phase space of protons not via perturbations to the trajectory, but in the history of the amplitude of the wavefield $\mathcal{E}(x,t)$ at the last point of resonance, which is calculated as follows.

\begin{figure}[H]
\begin{center}
\includegraphics[]{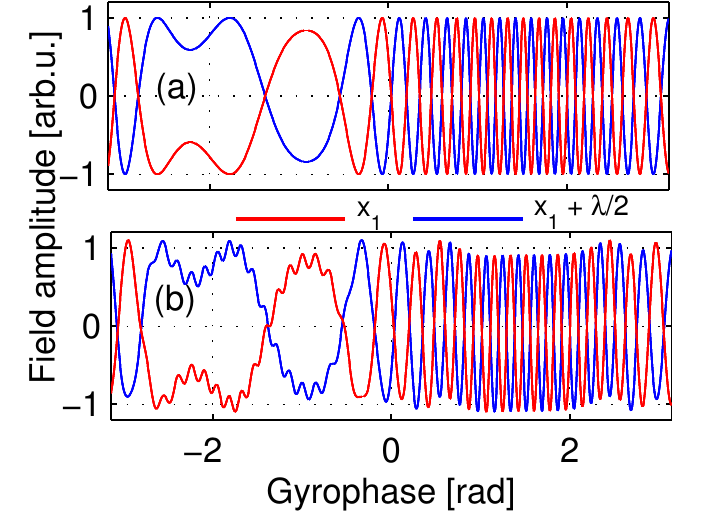}
\begin{small}
\caption{\label{fig:5a}Panel (a): Normalized wave amplitude seen by protons at resonance with the dominant backward wave, plotted as a function of gyrophase for two regions in configuration space separated by half the wavelength of the dominant wave (see text). Panel (b): As panel (a), for the sum of the backward propagating wavemode and the relatively weak forward wavemode.}
\end{small}
\end{center}
\end{figure}

The components of parallel and perpendicular velocities of a single proton gyro-orbit projected onto $v_x$ can be written, referring back to (\ref{eqn:falpha}), in the form
\begin{equation}
\label{eqn:vx}
v_x(t) = u\cos\theta - v_r(t)\sin\theta
\end{equation}
Here $\theta$ is the angle between the background magnetic field and the $x$-direction, and $v_{r}(t) = v_{r}\sin(\alpha(t))$, where the gyrophase is $\alpha(t) = \Omega_pt + \alpha_0$, and $\alpha_0$ is an initial condition. We can calculate the time of resonance $t_R$ for this proton by noting that at resonance $\omega/k = v_x$, so that
\begin{equation}
\label{eqn:tres}
v_x(t_R) = \omega/k = u\cos\theta - v_r\sin(\alpha(t_R))\sin\theta
\end{equation}
The position $x(t)$ of the proton is given by
\begin{equation}
x(t) = tu\cos\theta + v_r\cos(\alpha(t))\sin\theta/\Omega_{p} + x_0 
\end{equation}
where $x_0$ is the initial position of the proton. We eliminate the dependence on the initial condition $x_0$ by calculating the distance between the position of resonance $x_R$ and the position at time $t$,
\begin{equation}
 x(t) - x_R = (t-t_R)u\cos\theta + v_r\sin\theta(\cos(\alpha(t)) - \cos(\alpha(t_R)))/\Omega_{p}
\label{eqn:xminusxr}
\end{equation}
The normalized amplitude of the wave at the point of resonance is 
\begin{equation}
 \mathcal{E}(x,t) = \sin(kx_R(\alpha_0,x,t) - \omega t_R(\alpha_0))
\label{eqn:curlye}
\end{equation}
We now use (\ref{eqn:xminusxr}) and (\ref{eqn:curlye}) to generate a phase space diagram for protons distributed across a continuum of initial gyrophase. This is shown in Figure \ref{fig:5a}, which plots the normalized wave field $\mathcal{E}(x_R,t_R)$ as a function of proton gyrophase $\alpha(t)$ at points $(t_1,x_1)$ and $(t_1,x_1+\lambda/2)$. Figure \ref{fig:5a}(a) shows how protons initially uniformly distributed in gyrophase resonate with the dominant backward wave at $\omega \simeq 6\omega_{LH},k=-2\omega_{pe}/c$. This method recreates the large scale structure in Figure \ref{fig:VspaceAndVparaVsgyrophase}(b). The fine scale structure can be attributed in part to the interaction of the protons with the relatively weak forward travelling wave, as shown in Figure \ref{fig:5a}(b) which combines the effects of both waves. Here $\mathcal{E}(x,t) = sin(k_bx_{Rb} - \omega_b t_{Rb}) + 0.1sin(k_fx_{Rf} - \omega_f t_{Rf})$ is plotted at points $(t_1,x_1)$ and $(t_1,x_1+\lambda/2)$ against the gyrophase of the protons, where $(x_b,t_b)$ and $(x_f,t_f)$ are the points of resonance for the backward propagating wave $(\omega_b,k_b)$ and the weaker forward propagating wave $(\omega_f,k_f)$ respectively. In this case we take $\omega_f/k_f$ to equal the maximum in proton $v_x$.

\section{Conclusions}

We have studied a collective instability arising from spatially localised energetic ion populations that can arise in tokamaks, and the phenomenology of the resultant energy transfer to the thermal background plasma, notably the electrons. We employ a self-consistent relativistic electromagnetic 1D3V particle-in-cell code to evolve minority energetic protons and background thermal electrons and deuterons. The 1D3V PIC code provides 3D velocity and electromagnetic field vectors along one spatial direction and time. The single spatial direction restricts the propagation direction of waves to an oblique angle to the magnetic field. Plasma parameters are chosen to approximate where possible those at the outer mid-plane edge inside the last closed flux surface in a large tokamak plasma. The velocity distribution of confined fusion products is modelled by an antiparallel travelling ring beam, motivated by observations and interpretation of ion cyclotron emission in JET and TFTR. 

In using a PIC code, the initial perturbation is seeded by thermal noise that has broad spectral content, whereas analytical studies rely on eikonal approaches.  Computation of the proton fluctuation energy is valuable in demonstrating the existence of structured early time perturbations to the ensemble of particles, which are otherwise masked by noise. The initial perturbation to the proton distribution function generates a collective instability, resulting in growth of waves in which the electric field is substantially larger than the magnetic: predominantly electrostatic waves in the lower hybrid range of frequencies. Snapshots of the proton $v_{\parallel}$ distribution function enable us to link the multiharmonic mode structure of the electric field to oscillations in proton $v_{\parallel}$ as a function of gyrophase.

Fast Fourier transforms of the electric field along the simulation domain before the linear phase, and during its early and late stages, capture proton-excited $\omega,k$ structure. These waves drive electron acceleration through Landau damping. Directional asymmetry of the dominant excited wavemodes, combined with the Landau damping process, determines the direction of overall electron acceleration. The backward propagating dominant wavemode accelerates the electrons, generating a suprathermal tail, whereas the forward propagating wavemode interacts only weakly with the electrons.

We have thus provided evidence, by direct numerical simulation, for electron current drive from lower hybrid waves excited by confined fusion products in a tokamak-relevant parameter regime. This is the first time a fully kinetic electromagnetic PIC code has been used to model an instability produced by velocity space population inversion of confined fusion products that results in electron current. These results provide a key building block for designing future alpha channelling scenarios, which may open new regimes of lower hybrid current drive where external fields and local fusion product population inversions together channel energy into electrons. It also provides an interesting instance of self organisation \cite{ref:dendy2007ppcf} in plasmas.

\section{Acknowledgements}
This work was partly funded by the UK Engineering and Physics Sciences Research Council under grant EP/G003955 and by The European Communities under the contract of association between Euratom and CCFE. The views and opinions expressed herein do not necessarily represent those of the European Communities. The authors would like to thanks Christopher Brady and the epoch development team for their help with the code.

\section{References}

\nocite{ref:putvinski1998nuclfusion}

\begin{small}
\bibliographystyle{unsrt}
\bibliography{bibliography4}
\end{small}

\end{document}